\documentclass[]{elsarticle}

\usepackage{hyperref}
\usepackage{float}
\usepackage{graphicx}

\usepackage{booktabs}

\usepackage{tikz}
\usetikzlibrary{arrows.meta, arrows, calc, decorations.pathmorphing, shapes.geometric, positioning, shapes, fit, backgrounds, snakes}

\journal{Journal of \LaTeX\ Templates}

\begin{document}

\begin{frontmatter}

\title{Oscillatory Signatures of Parkinson's Disease: Central and Parietal EEG Alterations Across Multiple Frequency Bands}

\author[Australian National University]{Artem Lensky}

\ead{artem.lenskiy@anu.edu.au}

\begin{abstract}
This study investigates EEG as a potential early biomarker by applying deep learning techniques to resting-state EEG recordings from 31 subjects (15 with PD and 16 healthy controls). EEG signals underwent preprocessing to remove tremor artifacts before classification with CNNs using wavelet-based electrode triplet images. Our analysis across different brain regions and frequency bands showed distinct spatial-spectral patterns of PD-related neural oscillations. We identified high classification accuracy (76\%) using central electrodes (C3, Cz, C4) with full-spectrum 0.4-62.4 Hz analysis and 74\% accuracy in right parietal regions (P8, CP6, P4) with 10-second windows. Bilateral centro-parietal regions showed strong performance (67\%) in the theta band (4.0-7.79 Hz), while multiple areas demonstrated some sensitivity (65\%) in the alpha band (7.8-15.59 Hz). We also observed a distinctive topographical pattern of gamma band (40-62.4 Hz) alterations specifically localized to central-parietal regions, which remained consistent across different temporal windows. In particular, we observed pronounced right-hemisphere involvement across several frequency bands. Unlike previous studies that achieved higher accuracies by potentially including tremor artifacts, our approach isolates genuine neurophysiological alterations in cortical activity. These findings suggest that specific EEG-based oscillatory patterns, especially in central and parietal regions and across multiple frequency bands, may provide diagnostic information for PD, potentially before the onset of motor symptoms.
\end{abstract}

\end{frontmatter}

\section{Introduction}
Parkinson's disease (PD) is a progressive neurodegenerative disorder characterized primarily by motor symptoms including resting tremor, rigidity, bradykinesia, and postural instability \cite{Kalia2015}. PD is associated with the loss of dopaminergic neurons in the substantia nigra, resulting in dopamine deficiency \cite{Jankovic2008}. Despite decades of research, current diagnostic approaches still rely heavily on clinical observation of motor symptoms and response to dopaminergic medication—a significant limitation given that these manifestations typically appear only after 60-80\% of substantia nigra neurons have already degenerated \cite{Noyce2016}.

This substantial delay between neuronal loss and clinical diagnosis represents a critical window during which potential neuroprotective interventions might be most effective—if reliable early biomarkers could be identified. Non-motor symptoms, including REM sleep behavior disorder \cite{Breen2014}, hyposmia \cite{Doty2012}, constipation, and subtle cognitive changes \cite{Darweesh2016} often precede classical motor symptoms by years or even decades. However, attempts to develop screening protocols based on these prodromal features have yielded disappointing results. A recent five-year prospective study by Johnson et al. found that even combinations of three prodromal markers achieved positive predictive values of only 13.6\%, highlighting the urgent need for more objective physiological markers.

Electroencephalography (EEG) offers particular promise as a potential biomarker. Unlike neuroimaging techniques requiring expensive equipment and specialized facilities, EEG combines non-invasiveness with relatively low cost, portability, and widespread availability in clinical settings. Critically, EEG directly measures neuronal activity with millisecond temporal resolution, potentially capturing subtle alterations in neural synchronization patterns that may reflect early pathological processes \cite{Cohen2014}.

Traditional spectral analyses of EEG in PD have revealed somewhat inconsistent patterns across studies. While several groups have reported a general "slowing" of EEG activity—characterized by increased power in delta (0.4-4 Hz) and theta (4-8 Hz) bands with concurrent reductions in beta (13-30 Hz) and gamma ($>$30 Hz) power \cite{Li2020, Morita2009}—these findings have not been universally replicated. Notably, Serizawa et al. demonstrated that spectral alterations vary significantly with disease progression and medication state, finding increased rather than decreased beta activity in early-stage, unmedicated patients. Such contradictions likely stem partly from inconsistent electrode referencing schemes and normalization methods, as demonstrated when reanalyzing published datasets with standardized preprocessing pipelines.

The application of advanced computational approaches to EEG analysis in PD has gained momentum in recent years. A comprehensive systematic review by Maitín et al. \cite{Maitin2020} examined nine studies utilizing machine learning for PD classification based on EEG, finding accuracies ranging from 62\% to an impressive 99.62\%. However, critical evaluation of these studies reveals substantial methodological concerns. The highest reported accuracies (>95\%) typically emerged from studies with significant methodological limitations, including small sample sizes (often <20 subjects per group), inadequate cross-validation procedures, or insufficient separation between training and testing data. Furthermore, proper correction for multiple comparisons was frequently absent, raising concerns about potential false positive findings.

A particularly challenging aspect of EEG analysis in PD concerns artifact management \cite{Sazgar2019}. Unlike in other neurological conditions, PD patients exhibit pathological movements—including tremor at 4-6 Hz—that produce electrical signals spatially and temporally overlapping with neural oscillations of interest. Standard preprocessing approaches often fail to adequately distinguish these movement artifacts from genuine brain activity. For instance, blind source separation techniques like Independent Component Analysis (ICA) struggle with this separation because tremor-related components share spatial and temporal characteristics with cortical signals \cite{Weyhenmeyer2014, Whitham2007}. Previous attempts to address this challenge have often resulted in overly aggressive filtering that likely removes disease-relevant neural activity along with artifacts.

The spatial localization of PD-related EEG abnormalities presents another substantial challenge. Yet emerging evidence suggests significant involvement of parietal and occipital cortices in PD pathophysiology, with Tessitore et al. demonstrating altered resting-state functional connectivity in posterior cortical networks that correlates with specific cognitive symptoms \cite{Tessitore2012}. The reliance on predetermined regions of interest in most previous studies has potentially obscured important spatial patterns of EEG abnormalities across the cortex.

Particularly striking is the contrast between findings from Swann et al. \cite{Swann2015} and George et al. \cite{George2013} regarding beta oscillations in PD. While Swann reported elevated beta synchrony across sensorimotor cortex that correlated with motor symptom severity, George observed decreased cortical beta coherence following dopaminergic medication—a seemingly paradoxical finding given that medication improves motor symptoms. These contradictory results likely reflect methodological differences in coherence calculation and reference electrode selection, but also suggest a more complex, non-linear relationship between oscillatory activity and disease manifestations than previously recognized.

The field has lacked a comprehensive analysis of how PD affects different frequency bands across various brain regions. Most studies have examined either global measures (averaged across all electrodes) or preselected subsets of electrodes/frequencies based on a priori hypotheses. This selective approach risks missing unexpected spatial-spectral patterns that might provide even stronger diagnostic markers. The few studies attempting broader spatial analyses have typically employed inadequate statistical controls for multiple comparisons, raising the likelihood of chance findings.

The present study addresses these gaps by systematically evaluating EEG activity across multiple brain regions and frequency bands without preconceived notions about which combinations should prove most discriminative. By transforming EEG signals into wavelet-based images and grouping spatially adjacent electrodes, the approach captures local patterns of neural synchronization while addressing the limitations of traditional frequency analysis methods. This methodology enables comprehensive mapping of PD-related EEG alterations across the entire cortex and frequency spectrum.

Specifically, the study addresses three primary objectives:
\begin{enumerate}
    \item To develop and implement a robust preprocessing pipeline specifically designed to address the unique challenges of separating PD-related movement artifacts from neural oscillations, incorporating advanced artifact detection algorithms with minimal impact on disease-relevant signals.
    
    \item To systematically evaluate the discriminative power of different brain regions (electrode groups) and frequency bands without the regional biases evident in previous research, with particular emphasis on central and parietal regions and frequency bands (theta and alpha) that may provide the most discriminative information.
    
    \item To identify specific spatial-spectral patterns yielding optimal classification accuracy, with particular attention to central and parietal regions and their frequency-specific alterations that might serve as novel neurophysiological biomarkers for PD.
\end{enumerate}

The findings contribute to a more nuanced understanding of PD's impact on cortical oscillatory activity while offering potential EEG-based biomarkers that could eventually enable earlier diagnosis. The spatial-spectral specificity of these markers may also provide insights into the neurophysiological mechanisms underlying both motor and non-motor symptoms of PD. Understanding these patterns becomes increasingly important as research moves toward detecting prodromal stages of PD, when neuroprotective interventions might significantly alter disease progression. Nevertheless, the cross-sectional design employed here represents a first step, with longitudinal studies ultimately needed to validate these markers' utility for early detection.

\section{Methods}

\subsection{Dataset}
For this study, we utilized the openly available resting-state EEG dataset from the University of San Diego, specifically the "Rest EEG in Parkinson's Disease" dataset (OpenNeuro, ds002778, version 1.0.5) \cite{Jackson2019, Swann2015, George2013}. The dataset consists of resting-state EEG recordings from 31 participants: 15 individuals with PD and 16 healthy control subjects.

The EEG data was recorded using a 32-channel BioSemi ActiveTwo system at a sampling rate of 512 Hz. Recordings were conducted with participants in a resting state with their eyes open. Each participant underwent approximately 3 minutes of continuous EEG recording. The electrode placement followed the standard 10-20 international system.

It is important to note that the dataset has certain limitations acknowledged by the original data collectors. The Unified Parkinson's Disease Rating Scale (UPDRS) assessments were conducted by laboratory personnel who had completed online training rather than by board-certified neurologists, which may introduce some uncertainty in clinical evaluations.

Given the heterogeneous presentation of PD across patients and the relatively small sample size, we employed a cross-validation approach to maximize the statistical validity of our findings while acknowledging the limitations inherent to the dataset size.

\begin{figure}[H]
    \centering
    \begin{tikzpicture}[scale=4]
        \definecolor{myblue}{RGB}{0,113,188}
        \definecolor{myorange}{RGB}{216,82,24}
        \definecolor{myyellow}{RGB}{236,176,31}
        \definecolor{mypurple}{RGB}{125,46,141}
        \definecolor{mygreen}{RGB}{118,171,47}
        
        \draw[thick] (0,0) circle (1);
        
        \draw[gray, thin] (-1,0) -- (1,0);
        \draw[gray, thin] (0,-1) -- (0,1);
        
        \draw[thick] (-1,0) .. controls (-1.15,-0.1) and (-1.15,0.1) .. (-1,0.2);
        \draw[thick] (1,0) .. controls (1.15,-0.1) and (1.15,0.1) .. (1,0.2);
        
        \filldraw[fill=myblue, opacity=0.4, draw=myblue, thick] (-0.45,-0.05) rectangle (0.45,0.05);
        
        \filldraw[fill=myorange, opacity=0.4, draw=myorange, thick] (-0.7,-0.5) -- (-0.6,-0.3) -- (-0.4,-0.5) -- cycle;
        
        \filldraw[fill=myorange, opacity=0.4, draw=myorange, thick] (0.7,-0.5) -- (0.6,-0.3) -- (0.4,-0.5) -- cycle;
        
        \filldraw[fill=myyellow, opacity=0.1, draw=myyellow, thick] (-0.3,-0.75) -- (0,-0.5) -- (0.3,-0.75) -- cycle;
        
        \filldraw[fill=myorange, opacity=0.1, draw=myorange, thick] (-0.6,-0.3) -- (-0.4,0) -- (-0.2,-0.3) -- cycle;
        
        \filldraw[fill=myorange, opacity=0.1, draw=myorange, thick] (0.6,-0.3) -- (0.4,0) -- (0.2,-0.3) -- cycle;
        
        
        
        
        \filldraw[fill=mypurple!40, draw=black] (-0.3,0.75) circle (0.07) node[anchor=center] {\small{\texttt{AF3}}};
        \filldraw[fill=mypurple!40, draw=black] (0.3,0.75) circle (0.07) node[anchor=center] {\small{\texttt{AF4}}};
        
        \filldraw[fill=myblue!40, draw=black] (-0.7,0.5) circle (0.07) node[anchor=center] {\small{\texttt{F7}}};
        \filldraw[fill=myblue!40, draw=black] (-0.4,0.5) circle (0.07) node[anchor=center] {\small{\texttt{F3}}};
        \filldraw[fill=myblue!40, draw=black] (0,0.5) circle (0.07) node[anchor=center] {\small{\texttt{Fz}}};
        \filldraw[fill=myblue!40, draw=black] (0.4,0.5) circle (0.07) node[anchor=center] {\small{\texttt{F4}}};
        \filldraw[fill=myblue!40, draw=black] (0.7,0.5) circle (0.07) node[anchor=center] {\small{\texttt{F8}}};
        
        \filldraw[fill=myblue!20, draw=black] (-0.6,0.3) circle (0.07) node[anchor=center] {\small{\texttt{FC5}}};
        \filldraw[fill=myblue!20, draw=black] (-0.2,0.3) circle (0.07) node[anchor=center] {\small{\texttt{FC1}}};
        \filldraw[fill=myblue!20, draw=black] (0.2,0.3) circle (0.07) node[anchor=center] {\small{\texttt{FC2}}};
        \filldraw[fill=myblue!20, draw=black] (0.6,0.3) circle (0.07) node[anchor=center] {\small{\texttt{FC6}}};
        
        \filldraw[fill=gray!40, draw=black] (-0.8,0) circle (0.07) node[anchor=center] {\small{\texttt{T7}}};
        \filldraw[fill=gray!20, draw=black] (-0.4,0) circle (0.07) node[anchor=center] {\small{\texttt{C3}}};
        \filldraw[fill=gray!20, draw=black] (0,0) circle (0.07) node[anchor=center] {\small{\texttt{Cz}}};
        \filldraw[fill=gray!20, draw=black] (0.4,0) circle (0.07) node[anchor=center] {\small{\texttt{C4}}};
        \filldraw[fill=gray!40, draw=black] (0.8,0) circle (0.07) node[anchor=center] {\small{\texttt{T8}}};
        
        \filldraw[fill=myorange!30, draw=black] (-0.6,-0.3) circle (0.07) node[anchor=center] {\small{\texttt{CP5}}};
        \filldraw[fill=myorange!30, draw=black] (-0.2,-0.3) circle (0.07) node[anchor=center] {\small{\texttt{CP1}}};
        \filldraw[fill=myorange!30, draw=black] (0.2,-0.3) circle (0.07) node[anchor=center] {\small{\texttt{CP2}}};
        \filldraw[fill=myorange!30, draw=black] (0.6,-0.3) circle (0.07) node[anchor=center] {\small{\texttt{CP6}}};
        
        \filldraw[fill=myorange, draw=black] (-0.7,-0.5) circle (0.07) node[anchor=center] {\small{\texttt{P7}}};
        \filldraw[fill=myorange, draw=black] (-0.4,-0.5) circle (0.07) node[anchor=center] {\small{\texttt{P3}}};
        \filldraw[fill=myorange, draw=black] (0,-0.5) circle (0.07) node[anchor=center] {\small{\texttt{Pz}}};
        \filldraw[fill=myorange, draw=black] (0.4,-0.5) circle (0.07) node[anchor=center] {\small{\texttt{P4}}};
        \filldraw[fill=myorange, draw=black] (0.7,-0.5) circle (0.07) node[anchor=center] {\small{\texttt{P8}}};
        
        \filldraw[fill=myyellow!50, draw=black] (-0.3,-0.75) circle (0.07) node[anchor=center] {\small{\texttt{PO3}}};
        \filldraw[fill=myyellow!50, draw=black] (0.3,-0.75) circle (0.07) node[anchor=center] {\small{\texttt{PO4}}};
        
        \filldraw[fill=yellow!20, draw=black] (-0.25,-0.9) circle (0.07) node[anchor=center] {\small{\texttt{O1}}};
        \filldraw[fill=yellow!20, draw=black] (0,-0.9) circle (0.07) node[anchor=center] {\small{\texttt{Oz}}};
        \filldraw[fill=yellow!20, draw=black] (0.25,-0.9) circle (0.07) node[anchor=center] {\small{\texttt{O2}}};
        
        \node[right] at (1.05, 0.9) {\textbf{Regions:}};
        \filldraw[fill=mypurple!40, draw=black] (1.25, 0.8) circle (0.03) node[right=3pt] {\texttt{AF}};
        \filldraw[fill=myblue!40, draw=black] (1.25, 0.7) circle (0.03) node[right=3pt] {\texttt{F}};
        \filldraw[fill=myblue!20, draw=black] (1.25, 0.6) circle (0.03) node[right=3pt] {\texttt{FC}};
        \filldraw[fill=gray!20, draw=black] (1.25, 0.5) circle (0.03) node[right=3pt] {\texttt{C}};
        \filldraw[fill=gray!40, draw=black] (1.25, 0.4) circle (0.03) node[right=3pt] {\texttt{T}};
        \filldraw[fill=myorange!30, draw=black] (1.25, 0.3) circle (0.03) node[right=3pt] {\texttt{CP}};
        \filldraw[fill=myorange, draw=black] (1.25, 0.2) circle (0.03) node[right=3pt] {\texttt{P}};
        \filldraw[fill=myyellow!50, draw=black] (1.25, 0.1) circle (0.03) node[right=3pt] {\texttt{PO}};
        \filldraw[fill=yellow!20, draw=black] (1.25, 0.0) circle (0.03) node[right=3pt] {\texttt{O}};
        
        
    \end{tikzpicture}
    \caption{Electrode placement according to the 10-20 international system showing the 32 electrodes used in this study. The light and dark shaded areas indicate electrode triplets with classification accuracy above 65\% and above 70\% correspondingly.}
    \label{fig:electrode_placement}
\end{figure}
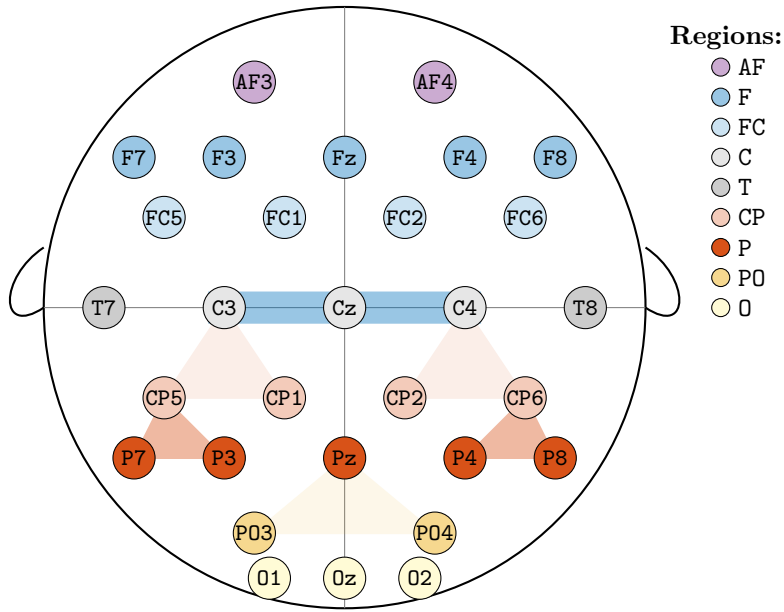

\subsection{EEG Preprocessing}
The raw EEG data underwent a preprocessing pipeline using MATLAB EEGLAB toolbox \cite{Appelhoff2019, Pernet2019} to remove artifacts and prepare the signals for wavelet transformation and subsequent analysis. The preprocessing steps were as follows:

The EEG data was downsampled from 512 Hz to 128 Hz. This step reduced computational load while maintaining sufficient temporal resolution for frequency analysis up to 64 Hz.
    
Then, a finite impulse response (FIR) bandpass filter to remove frequency components below 0.4 Hz and above 63.5 Hz. This eliminated slow drifts, DC offset, and high-frequency noise while preserving the frequency bands of interest (delta, theta, alpha, beta, and gamma).
    
Lastly, we applied Independent Component Analysis to decompose the EEG signals into independent components and identify components associated with artifacts. Components classified as non-brain sources (e.g., eye movements, muscle activity) were removed, and the signals were reconstructed.
\begin{figure}
    \centering
    \includegraphics[width=0.7\linewidth]{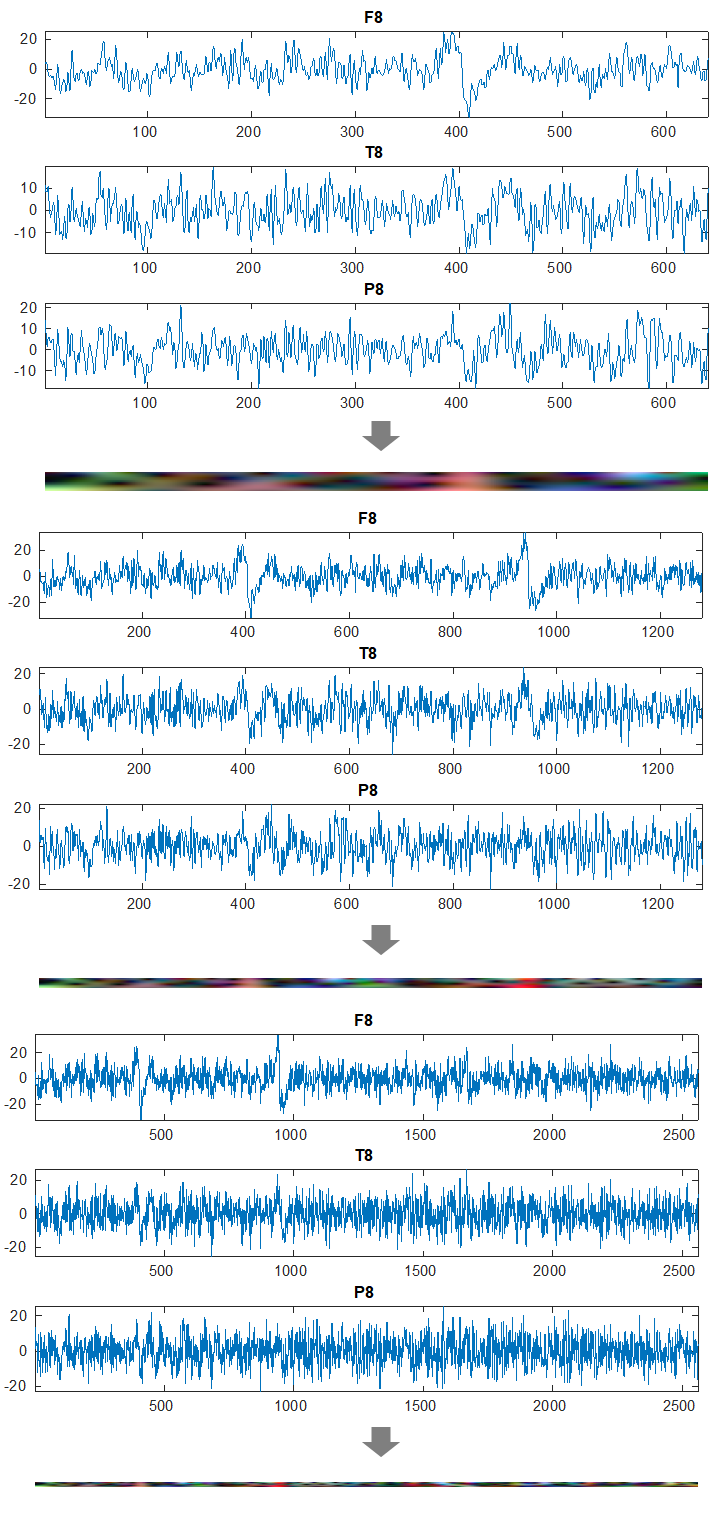}
    \caption{Transformation of raw EEG signals RGB images. The signals are segmented into different window lengths (5, 10, and 20 seconds) followed by wavelet transformation to create 2D time-frequency representations, form which RGB images are formed by combining triplets of electrode channels.}
    \label{fig:preprocessing_pipeline}
\end{figure}

\subsection{Data Preparation}
After preprocessing, the continuous EEG recordings were segmented into overlapping windows of three different durations: 5 seconds, 10 seconds, and 20 seconds. This windowing approach served multiple purposes:

\begin{enumerate}
    \item Increase effective sample size by creating multiple examples from each participant's recording.
    \item Compare the classification performance across different temporal scales and assess the stability and consistency of the results.
\end{enumerate}

For each window, we applied continuous wavelet transform (CWT) to convert the one-dimensional time series into two-dimensional time-frequency representations.  The resulting wavelet coefficients were analyzed across six frequency bands: delta (0.4-3.99 Hz), theta (4.0-7.79 Hz), alpha (7.8-15.59 Hz), beta1 (15.6-31.19 Hz), beta2 (31.2-39.99 Hz), and gamma (40-62.4 Hz).

To prepare the data for input to the convolutional neural network (CNN), we organized the EEG channels into 20 groups of three spatially adjacent electrodes based on their proximity on the scalp. Figure \ref{fig:electrode_placement} shows 6 triplets (groups of 3 channels) showing the highest classification accuracy. Altogether these were the 20 electrode triplets split into 4 groups:

\begin{table}[H]
    \centering
    \begin{tabular}{c|l l}
        \toprule
        \textbf{Group} & \textbf{Electrode Triplets} & \textbf{Brain Regions} \\
        \midrule
        1 & \small\texttt{\{AF3, Fz, AF4\}}, \texttt{\{FC1, Fz, FC2\}}, & Frontal and Fronto-Central \\
          & \small{\texttt{\{C3, Cz, C4\}}},  \small\texttt{\{FC5, F3, FC1\}}, &  Regions \\
          & \texttt{\{FC6, F4, FC2\}} &  \\
        \hline
        2 & \small\texttt{\{CP1, Pz, CP2\}}, \texttt{\{P3, Pz, P4\}},  & Central-Parietal Regions \\
          & \small{\texttt{\{PO3, Pz, PO4\}}}, \small\texttt{\{CP5, C3, CP1\}}, &  \\
          & \texttt{\{CP6, C4, CP2\}} & \\
        \hline
        3 & \small\texttt{\{O1, Oz, O2\}}, \texttt{\{PO3, O1, Oz\}}, & Parieto-Occipital Regions \\
          & \small{\texttt{\{PO4, O2, Oz\}}}, \small\texttt{\{P7, CP5, P3\}}, & \\
          & \texttt{\{P8, CP6, P4\}} &  \\
        \hline
        4 & \small\texttt{\{F7, T7, P7\}}, \texttt{\{F3, C3, P3\}}, & Longitudinal/Cross-Regional \\
          & \small{\texttt{\{Cz, Pz, Oz\}}}, \small\texttt{\{F4, C4, P4\}}, & Connections \\ 
          & \texttt{\{F8, T8, P8\}} & \\
        \bottomrule
    \end{tabular}
    \caption{Organization of electrode triplets into four major functional groups based on their spatial distribution across the scalp.}
    \label{tab:electrode_groups}
\end{table}

\normalsize For each electrode triplet, the corresponding wavelet coefficient matrices were normalized, with 1\% of peak values removed to reduce the impact of outliers. These processed matrices were then combined to form RGB images, where each color channel (red, green, blue) represented the wavelet coefficients from one electrode in the triplet (Figure \ref{fig:preprocessing_pipeline}). This approach allowed us to leverage the power of convolutional neural networks designed for image processing while maintaining the spatial relationships between electrodes.

\subsection{Convolutional Neural Network Model}
We experimented with several convolutional neural network (CNN) including SqueezeNet, GoogleNetm Inception-v3, ResNet, to classify the wavelet-based RGB images as either PD or healthy control, but out of all tested models, a simple simple convolutional neural network (CNN) architecture achieved the highest performance and required substantially less computational resources. This choice also allowed us:

\begin{enumerate}
    \item To focus on extracting fundamental features that differentiate PD from healthy controls, demonstrating that EEG signals contain rich information about brain activity patterns even with basic model architectures.
    \item To address the relatively small dataset size, where models with fewer parameters are less prone to overfitting.
\end{enumerate}

\begin{figure}
    \centering
    \includegraphics[width=0.8\linewidth]{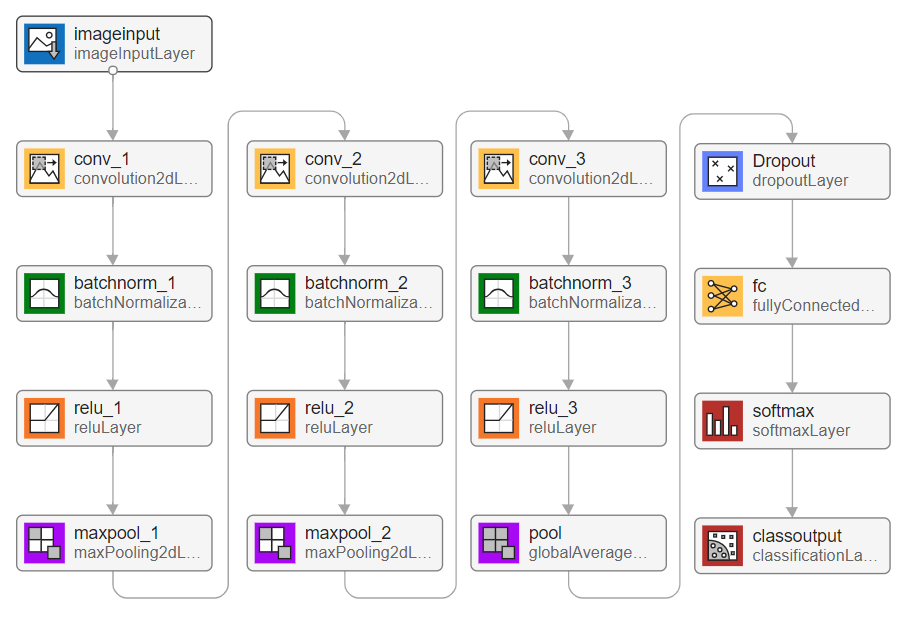}
    \caption{Architecture of the convolutional neural network used for classification. The model consists of three convolutional layers with batch normalization and ReLU activation, followed by max pooling, dropout for regularization, and a fully connected layer for binary classification (PD vs. healthy control). Different hyperparameters were optimized for each window length (5, 10, and 20 seconds) to maximize classification performance.}
    \label{fig:cnn_architecture}
\end{figure}

The CNN architecture, as illustrated in Figure \ref{fig:cnn_architecture}, consisted of the following components:

\begin{table}[H]
    \centering
    \begin{tabular}{c|l}
        \toprule
        \textbf{Layer Type} & \textbf{Description} \\
        \midrule
        Input Layer & Accepts RGB images created from wavelet coefficients\\ & of electrode triplets \\
        \midrule
        Convolutional Layers & Three convolutional layers (16, 32, 128 filters) \\
        & All using 5×5 kernel size with padding of 1 \\
        & Batch normalization after each convolution \\
        \midrule
        Activation Functions & ReLU applied after each convolutional layer \\
        \midrule
        Pooling Layers & Max pooling (4×4, stride 4) after first two conv. blocks \\
        & Global average pooling after the final conv. block \\
        \midrule
        Dropout Layer & Applied before the fully connected layer \\
        & Prevents overfitting by randomly deactivating neurons \\
        \midrule
        Fully Connected Layer & Binary classification output (PD vs. healthy control) \\
        \bottomrule
    \end{tabular}
    \caption{CNN architecture components used for classifying EEG-derived wavelet images.}
    \label{tab:cnn_architecture}
\end{table}

The entire project was implemented in MATLAB, including preprocessing, feature extraction, model training, and evaluation.

\subsection{Hyperparameter Optimization and Training Protocol}
For each window length (5, 10, and 20 seconds), we conducted a Bayesian hyperparameter search using only the all data to identify the optimal model configuration. The key hyperparameters that were optimized included: Initial Learning Rate, Dropout Rate, Momentum and Batch Size.

The optimal hyperparameters identified for each window length were:

\begin{table}[H]
    \centering
    \begin{tabular}{l|c|c|c}
        \toprule
        \textbf{Hyperparameter} & \textbf{5-second} & \textbf{10-second} & \textbf{20-second} \\
        \midrule
        Initial Learning Rate & 0.0034287 & 0.0065298 & 0.33049 \\
        Dropout Rate & 0.25291 & 0.37323 & 0.050525 \\
        Momentum & 0.11418 & 0.29409 & 0.58991 \\
        Batch Size & 47 & 37 & 29 \\
        \bottomrule
    \end{tabular}
    \caption{Optimal hyperparameters identified for CNN models with different window lengths.}
    \label{tab:hyperparameters}
\end{table}

Networks were trained using Stochastic Gradient Descent with Momentum (SGDM) optimizer for a maximum of 100 epochs. A piecewise learning rate schedule was employed with a drop period of 20 epochs and a drop factor of 0.5.

\subsection{Classification Procedure and Evaluation}
To evaluate the discriminative power of different brain regions and frequency bands, we trained and tested separate CNN models for each combination of electrode triplet and frequency band. We employed a leave-one-subject-out cross-validation approach to ensure robust assessment of classification accuracy while maximizing the use of our limited dataset. In this approach, data from one subject was used for testing while data from the remaining 30 subjects were used for training. This process was repeated for each subject, resulting in 31 separate models per electrode-frequency combination.

For each model, we computed classification accuracy as the primary performance metric, defined as the percentage of correctly classified instances (both PD and healthy controls) out of the total number of test instances. The classification error was calculated as 1 - (number of correct predictions / total number of predictions). Given the balanced nature of our dataset (15 PD, 16 controls), accuracy provides a fair assessment of classification performance.

\section{Results}

\subsection{Overall Classification Performance}
The overall classification accuracy varied substantially across different window lengths, electrode triplets, and frequency bands. Table \ref{tab:window_comparison} summarizes the mean classification accuracy across all electrode triplets for each window length using the full frequency spectrum (0.4-62.4 Hz).

\begin{table}[H]
    \centering
    \begin{tabular}{l|c}
        \toprule
        \textbf{Window Length} & \textbf{Mean Accuracy (All Electrodes)} \\
        \midrule
        5-second windows & 58.4\% $\pm$ 29.8\% \\
        10-second windows & 59.9\% $\pm$ 34.4\% \\
        20-second windows & 60.2\% $\pm$ 37.5\% \\
        \bottomrule
    \end{tabular}
    \caption{Mean classification accuracy across all electrode triplets for different window lengths using the full frequency spectrum. Note that the differences across window lengths are minimal, suggesting that PD-related EEG features are present at various temporal scales.}
    \label{tab:window_comparison}
\end{table}

\subsection{Regional Classification Accuracy}
Classification accuracy varied considerably across different electrode triplets. Table \ref{tab:high_acc_regions} presents the accuracies for the best-performing electrode triplets across all window lengths using the full frequency spectrum.

\begin{table}[H]
    \centering
    \begin{tabular}{l|c c c c}
        \toprule
        \textbf{Electrode Triplet} & \textbf{Group} & \textbf{5s Window} & \textbf{10s Window} & \textbf{20s Window} \\
        \midrule
        \texttt{\{C3, CZ, C4\}} & 1 & 66\% $\pm$ 28\% & 68\% $\pm$ 33\% & 76\% $\pm$ 32\% \\
        \texttt{\{CP5, C3, CP1\}} & 2 & 68\% $\pm$ 30\% & 70\% $\pm$ 35\% & 67\% $\pm$ 38\% \\
        \texttt{\{CP6, C4, CP2\}} & 2 & 70\% $\pm$ 30\% & 67\% $\pm$ 36\% & 68\% $\pm$ 38\% \\
        \texttt{\{P7, CP5, P3\}} & 3 & 61\% $\pm$ 23\% & 61\% $\pm$ 32\% & 73\% $\pm$ 32\% \\
        \texttt{\{P8, CP6, P4\}} & 3 & 69\% $\pm$ 27\% & 74\% $\pm$ 30\% & 73\% $\pm$ 34\% \\
        \texttt{\{F8, T8, P8\}} & 4 & 65\% $\pm$ 29\% & 63\% $\pm$ 37\% & 64\% $\pm$ 41\% \\
        \bottomrule
    \end{tabular}
    \caption{Classification accuracy for top-performing electrode triplets across different window lengths using the full frequency spectrum (0.4-62.4 Hz). Values represent mean accuracy $\pm$ standard deviation. The group number corresponds to the electrode groupings in Table \ref{tab:electrode_groups}.}
    \label{tab:high_acc_regions}
\end{table}

The central electrode triplet \texttt{\{C3, CZ, C4\}} from Group 1 achieved the highest overall accuracy of 76\% $\pm$ 32\% with 20-second windows, while showing 68\% $\pm$ 33\% accuracy with 10-second windows. The right parietal region \texttt{\{P8, CP6, P4\}} from Group 3 demonstrated particularly strong performance, reaching 74\% $\pm$ 30\% accuracy with 10-second windows and maintaining similar performance (73\% $\pm$ 34\%) with 20-second windows. These findings suggest that both central motor and right parietal regions contain robust discriminative information for PD classification, with central regions potentially benefiting more from longer temporal windows.

\begin{figure}
    \centering
    \includegraphics[width=0.5\linewidth]{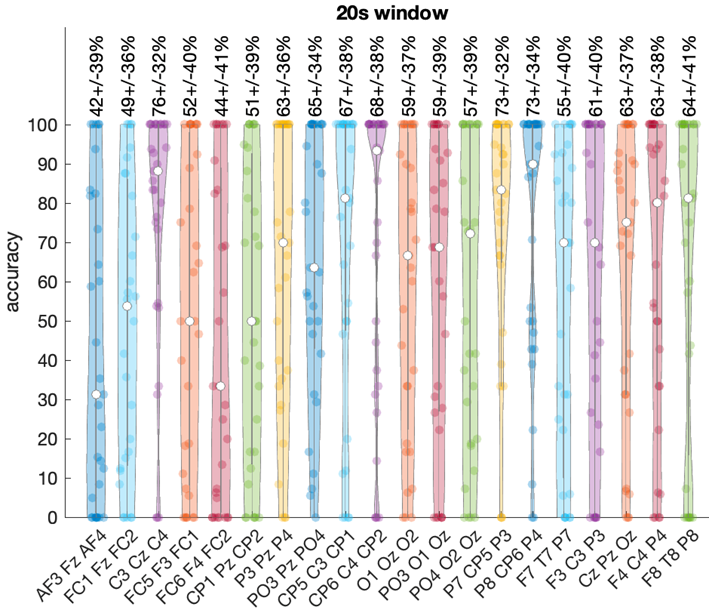}
    \caption{Violin plot showing the distribution of classification accuracy for each electrode triplet using 20-second windows and the full frequency spectrum.}
    \label{fig:violin_plots_20s}
\end{figure}

The left parietal region \texttt{\{P7, CP5, P3\}} showed a substantial improvement from 61\% $\pm$ 32\% with 10-second windows to 73\% $\pm$ 32\% with 20-second windows, suggesting that longer sampling periods may be important for capturing discriminative features in this region. Figure \ref{fig:violin_plots_20s} illustrates the variability in performance and highlights the superior accuracy of central and parietal regions. Note that because data from one subject was used for testing while data from the remaining 30 subjects were used for training, hence every violin plot contains exactly 30 points.

\subsection{Group-Level Performance}

Table \ref{tab:group_performance} shows the mean classification accuracy for each electrode group across different window lengths.

\begin{table}[H]
    \centering
    \begin{tabular}{l|c c c }
        \toprule
        \textbf{Electrode Group} & \textbf{5s Window} & \textbf{10s Window} & \textbf{20s Window} \\
        \midrule
        1: Frontal/Fronto-Central & 52.8\% $\pm$ 31.6\% & 53.4\% $\pm$ 36.0\% & 52.6\% $\pm$ 37.6\% \\
        2: Central-Parietal & 63.0\% $\pm$ 28.6\% & 64.2\% $\pm$ 33.8\% & 62.8\% $\pm$ 37.0\% \\
        3: Parieto-Occipital & 60.0\% $\pm$ 27.8\% & 62.6\% $\pm$ 32.4\% & 64.2\% $\pm$ 36.2\% \\
        4: Longitudinal Connections & 57.6\% $\pm$ 31.0\% & 59.2\% $\pm$ 35.4\% & 61.2\% $\pm$ 39.2\% \\
        \bottomrule
    \end{tabular}
    \caption{Mean classification accuracy for each electrode group across different window lengths using the full frequency spectrum. Values represent mean accuracy $\pm$ standard deviation across all electrode triplets within each group.}
    \label{tab:group_performance}
\end{table}

Group 3 (Parieto-Occipital) showed the highest mean accuracy with 20-second windows (64.2\% $\pm$ 36.2\%), followed by Group 2 (Central-Parietal) with 10-second windows (64.2\% $\pm$ 33.8\%). Group 1 (Frontal/Fronto-Central) showed the lowest mean accuracy across all window lengths, though still above chance level. However, it is worth noting that within Group 1, the central electrode triplet \texttt{\{C3, CZ, C4\}} significantly outperformed other triplets in the same group, highlighting the regional specificity of discriminative EEG patterns even within broader anatomical groupings.

\subsection{Frequency Band Analysis}
We analyzed the classification performance across six frequency bands: delta (0.4-3.99 Hz), theta (4.0-7.79 Hz), alpha (7.8-15.59 Hz), beta1 (15.6-31.19 Hz), beta2 (31.2-39.99 Hz), and gamma (40-62.4 Hz). Table \ref{tab:freq_band_acc} presents the classification accuracy for key electrode triplets across different frequency bands using 10-second windows.

\begin{table}[H]
    \centering
    \begin{tabular}{l|c| c c c c c c}
        \toprule
        \textbf{Electrode Triplet} & \textbf{Group} & \textbf{Delta} & \textbf{Theta} & \textbf{Alpha}\\
        \midrule
        $\texttt{\{C3, CZ, C4\}}$ & 1 & 58\% $\pm$ 29\% & 59\% $\pm$ 27\% & 65\% $\pm$ 29\%  \\
        $\texttt{\{CP5, C3, CP1\}}$ & 2 & 54\% $\pm$ 23\% & 67\% $\pm$ 30\% & 64\% $\pm$ 28\% \\
        $\texttt{\{CP6, C4, CP2\}}$ & 2 & 56\% $\pm$ 25\% & 67\% $\pm$ 31\% & 57\% $\pm$ 26\%  \\
        $\texttt{\{P7, CP5, P3\}}$ & 3 & 52\% $\pm$ 23\% & 59\% $\pm$ 27\% & 58\% $\pm$ 25\%  \\
        $\texttt{\{P8, CP6, P4\}}$ & 3 & 55\% $\pm$ 24\% & 60\% $\pm$ 28\% & 65\% $\pm$ 29\% \\
        $\texttt{\{PO3, PZ, PO4\}}$ & 2 & 52\% $\pm$ 23\% & 62\% $\pm$ 28\% & 57\% $\pm$ 25\%  \\
        $\texttt{\{F3, C3, P3\}}$ & 4 & 51\% $\pm$ 22\% & 59\% $\pm$ 27\% & 65\% $\pm$ 29\%  \\
        $\texttt{\{F8, T8, P8\}}$ & 4 & 58\% $\pm$ 26\% & 63\% $\pm$ 28\% & 65\% $\pm$ 29\%  \\
        $\texttt{\{P4, O4, PO4\}}$ & 3 & 51\% $\pm$ 22\% & 58\% $\pm$ 26\% & 57\% $\pm$ 25\% \\
        \bottomrule
    \end{tabular}
    \vspace{0.5cm}

    \begin{tabular}{l|c| c c c c c c}
        \toprule
        \textbf{Electrode Triplet} & \textbf{Group} & \textbf{Beta1} & \textbf{Beta2} & \textbf{Gamma} \\
        \midrule
        $\texttt{\{C3, CZ, C4\}}$   & 1 & 56\% $\pm$ 25\% & 49\% $\pm$ 21\% & 55\% $\pm$ 24\% \\
        $\texttt{\{CP5, C3, CP1\}}$ & 2 & 65\% $\pm$ 31\% & 50\% $\pm$ 23\% & 56\% $\pm$ 25\% \\
        $\texttt{\{CP6, C4, CP2\}}$ & 2 & 49\% $\pm$ 22\% & 54\% $\pm$ 24\% & 54\% $\pm$ 24\% \\
        $\texttt{\{P7, CP5, P3\}}$  & 3 & 49\% $\pm$ 21\% & 47\% $\pm$ 21\% & 51\% $\pm$ 23\% \\
        $\texttt{\{P8, CP6, P4\}}$  & 3 & 49\% $\pm$ 22\% & 44\% $\pm$ 19\% & 52\% $\pm$ 23\% \\
        $\texttt{\{PO3, PZ, PO4\}}$ & 2 & 48\% $\pm$ 21\% & 47\% $\pm$ 21\% & 47\% $\pm$ 20\% \\
        $\texttt{\{F3, C3, P3\}}$   & 4 & 49\% $\pm$ 22\% & 53\% $\pm$ 24\% & 54\% $\pm$ 24\% \\
        $\texttt{\{F8, T8, P8\}}$   & 4 & 57\% $\pm$ 26\% & 55\% $\pm$ 25\% & 52\% $\pm$ 23\% \\
        $\texttt{\{P4, O4, PO4\}}$  & 3 & 50\% $\pm$ 22\% & 54\% $\pm$ 24\% & 52\% $\pm$ 22\% \\
        \bottomrule
    \end{tabular}

    \caption{Classification accuracy for key electrode triplets across different frequency bands using 10-second windows. Values represent mean accuracy $\pm$ standard deviation. The group number corresponds to the electrode groupings in Table \ref{tab:electrode_groups}.}
    \label{tab:freq_band_acc}
\end{table}

Bilateral centro-parietal regions (\texttt{\{CP5, C3, CP1\}} and \texttt{\{CP6, C4, CP2\}}) from Group 2 showed strong performance (67\% $\pm$ 30-31\%) specifically in the theta band. The left centro-parietal region (\texttt{\{CP5, C3, CP1\}}) also achieved notable performance (65\% $\pm$ 31\%) in the beta1 band. Multiple regions reached approximately 65\% accuracy in the alpha band, including the central region (\texttt{\{C3, CZ, C4\}}), right parietal region (\texttt{\{P8, CP6, P4\}}), and two longitudinal connections (\texttt{\{F3, C3, P3\}} and \texttt{\{F8, T8, P8\}}). This widespread alpha band sensitivity across functionally distinct regions suggests that alpha oscillatory disruptions may represent a more global marker of PD-related cortical dysfunction.

The specificity of theta band sensitivity to bilateral centro-parietal regions is particularly noteworthy, as these regions are known to be involved in sensorimotor integration functions. The central-parietal region (\texttt{\{CP1, Pz, CP2\}}) from Group 2 showed moderate performance in the gamma band. Overall, the theta and alpha bands demonstrated the most consistent discriminative power across different regions, suggesting their particular relevance for PD classification.

\begin{figure}
    \centering
    \includegraphics[width=1\linewidth]{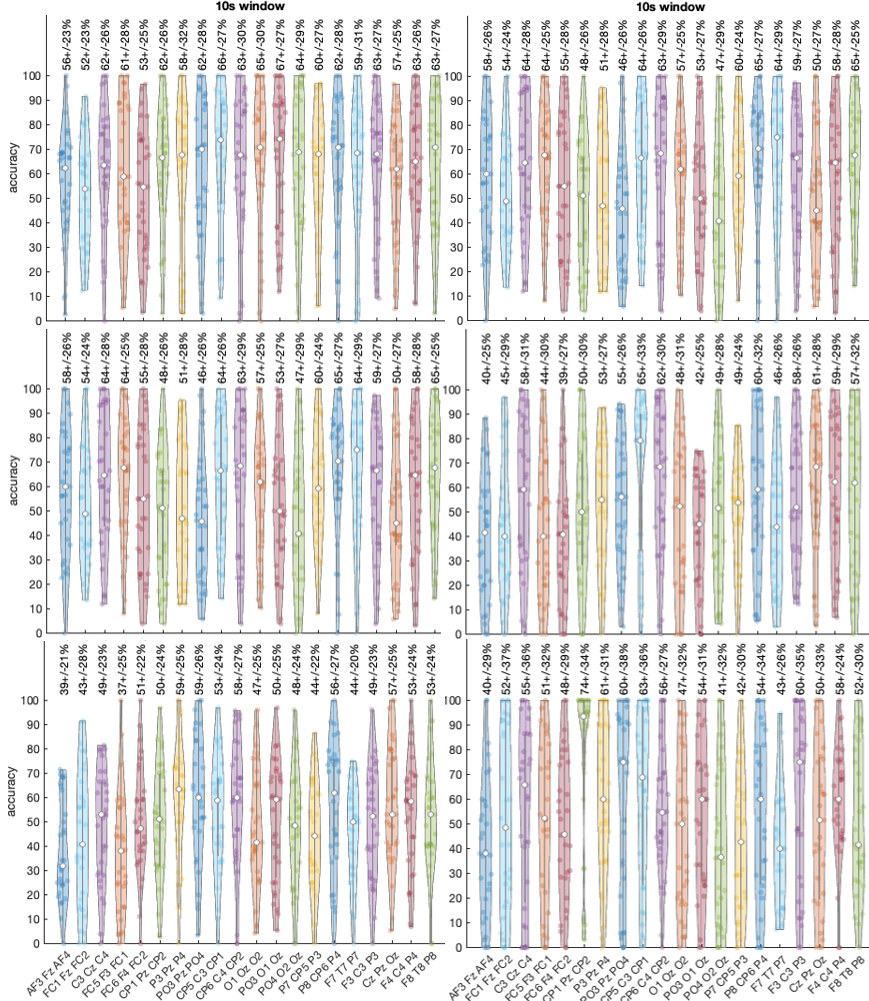}
    \caption{Violin plots showing the distribution of classification accuracy using 10-second windows for each electrode triplet across different frequency bands: delta (top-left), theta (top-right), alpha (mid-left), beta1 (mid-right), beta2 (bottom-left), gamma (bottom-right).}
    \label{fig:frequency_bands_violin}
\end{figure}

Figure \ref{fig:frequency_bands_violin} illustrates the distribution of classification accuracy across all electrode triplets for each frequency band using 10-second windows.

Table \ref{tab:group_freq_band_performance} shows the mean classification accuracy for each electrode group across different frequency bands using 10-second windows.

\begin{table}[H]
    \centering
    \begin{tabular}{l| c c c}
        \toprule
        \textbf{Electrode Group} & \textbf{Delta} & \textbf{Theta} & \textbf{Alpha} \\
        \midrule
        1: Frontal/Fronto-Central & 49.0\% $\pm$ 18.0\% & 62.0\% $\pm$ 26.0\% & 64.0\% $\pm$ 28.0\% \\
        2: Central-Parietal & 52.0\% $\pm$ 20.7\% & 63.7\% $\pm$ 28.3\% & 57.7\% $\pm$ 27.0\%  \\
        3: Parieto-Occipital & 50.5\% $\pm$ 25.0\% & 61.0\% $\pm$ 27.5\% & 62.5\% $\pm$ 25.5\% \\
        4: Longitudinal Connections & 53.5\% $\pm$ 20.5\% & 63.0\% $\pm$ 27.0\% & 62.0\% $\pm$ 26.0\% \\
        \bottomrule
    \end{tabular}
    \vspace{0.5cm}

    \begin{tabular}{l| c c c}
        \toprule
        \textbf{Electrode Group} & \textbf{Beta1} & \textbf{Beta2} & \textbf{Gamma} \\
        \midrule
        1: Frontal/Fronto-Central & 58.0\% $\pm$ 31.0\% & 49.0\% $\pm$ 23.0\% & 55.0\% $\pm$ 36.0\% \\
        2: Central-Parietal & 60.7\% $\pm$ 29.7\% & 56.7\% $\pm$ 25.7\% & 59.7\% $\pm$ 33.7\% \\
        3: Parieto-Occipital & 54.5\% $\pm$ 28.0\% & 50.0\% $\pm$ 24.5\% & 48.0\% $\pm$ 32.0\% \\
        4: Longitudinal Connections & 57.5\% $\pm$ 29.0\% & 51.0\% $\pm$ 23.5\% & 56.0\% $\pm$ 32.5\% \\
        \bottomrule
    \end{tabular}
    \caption{Mean classification accuracy for each electrode group across different frequency bands using 10-second windows. Values represent mean accuracy $\pm$ standard deviation across electrode triplets within each group.}
    \label{tab:group_freq_band_performance}
\end{table}

Each electrode group showed distinct patterns of frequency-specific discriminative power. Group 1 (Frontal/Fronto-Central) showed the highest accuracy in the alpha band (64.0\% $\pm$ 28.0\%). Group 2 (Central-Parietal) showed peak performance in the theta band (63.7\% $\pm$ 28.3\%). Group 3 (Parieto-Occipital) showed highest accuracy in the alpha band (62.5\% $\pm$ 25.5\%). Group 4 (Longitudinal Connections) showed highest accuracy in the theta band (63.0\% $\pm$ 27.0\%). This regional specificity of frequency band alterations suggests that PD affects different functional networks through distinct oscillatory mechanisms, which may reflect the heterogeneous pathophysiology of the disease.

\subsection{Topographical Patterns}
Topographical maps were generated to visualize the spatial distribution of classification accuracy across the scalp for each frequency band. It is important to note that the color scales vary across different topographical maps, with yellow representing different peak accuracy values in each map (ranging from approximately 56\% in some bands to 74\% in others). This calibration allows for better visualization of the spatial patterns within each frequency band. Despite these scale differences, consistent spatial patterns emerge across window lengths, indicating robust regional EEG alterations associated with PD.

\begin{figure}[H]
    \centering
    \includegraphics[width=1\textwidth]{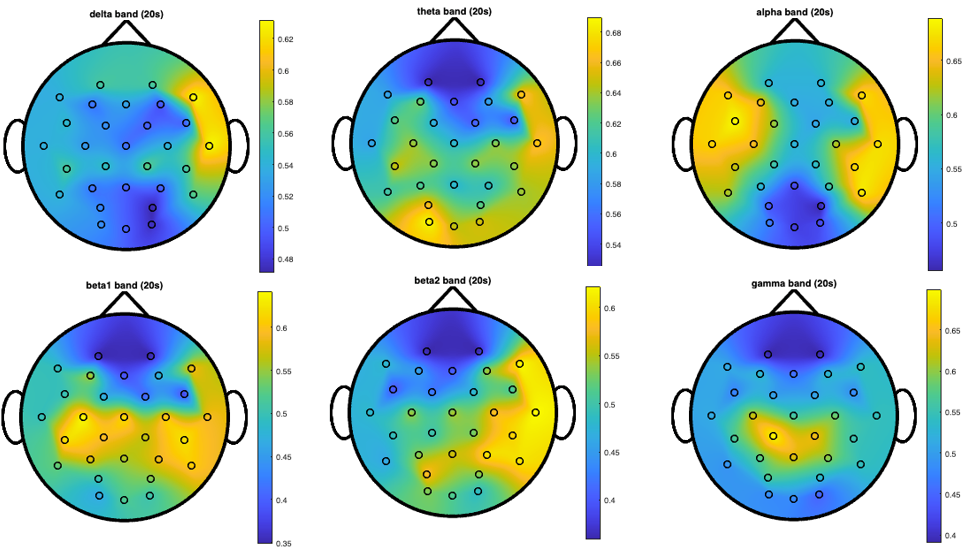}
    \caption{Topographical maps showing the spatial distribution of classification accuracy for different frequency bands using 20-second windows. Warmer colors (yellow) indicate regions with higher discriminative power, while cooler colors (blue) represent regions with lower discriminative power.}
    \label{fig:topography_all_bands}
\end{figure}

Figure \ref{fig:topography_all_bands} presents the topographical distribution of classification accuracy across different frequency bands using 20-second windows, showing distinct spatial patterns of discriminative power. 

\begin{figure}[H]
    \centering
    \includegraphics[width=0.9\textwidth]{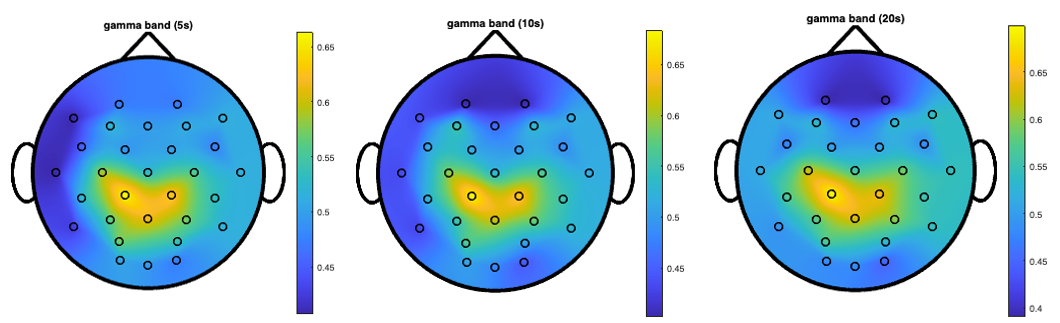}
    \caption{Comparison of gamma band (40-62.4 Hz) topographical maps across different window lengths (5, 10, and 20 seconds).}
    \label{fig:window_comparison_gamma}
\end{figure}

A particularly finding from our topographical analysis is the remarkable consistency of gamma band (40-62.4 Hz) spatial distribution across different window lengths (Figure \ref{fig:window_comparison_gamma}). This spatial specificity, combined with its temporal stability across different window lengths (5, 10, and 20 seconds), suggests that gamma band alterations reflect a fundamental aspect of PD pathophysiology rather than a transient feature, that is also supported by \cite{FOFFANI2022259}. While the central-parietal region (\texttt{\{CP1, Pz, CP2\}}) showed moderate performance in the gamma band overall, the distinct topographical pattern of gamma alterations compared to other frequency bands indicates that high-frequency neural synchronization in central-parietal regions may represent a specific marker of the disease. This contrasts with lower frequency bands (theta and alpha) which show more distributed patterns of discriminative power across multiple brain regions. 

Further, a consistent pattern across multiple frequency bands is the right-hemisphere emphasis in classification accuracy, particularly evident in the delta, beta2, and alpha bands. This lateralization of EEG alterations suggests asymmetric neurophysiological effects of PD on brain oscillatory activity, with more pronounced or detectable changes in right-hemisphere regions involved in visuospatial processing and sensorimotor integration.

\section{Discussion}

\subsection{Interpretation of Spatial-Spectral Patterns}

Our results demonstrate distinct spatial-spectral EEG patterns that differentiate PD patients from healthy controls, with several specific regions and frequency bands showing particularly strong discriminative power.

\subsubsection{Functional Group Analysis}

By organizing electrode triplets into four functional groups, we identified how PD differentially affects various brain networks:

Group 1 (Frontal/Fronto-Central regions) showed variability in classification performance, with the central electrode triplet \texttt{\{C3, CZ, C4\}} significantly outperforming other triplets in this group and reaching 76\% accuracy with 20-second windows. This finding aligns with PD's well-documented disruption of cortico-basal ganglia-thalamo-cortical motor circuits \cite{Kalia2015, Jankovic2008}. The central electrodes' high discriminative power likely reflects underlying pathophysiological changes in primary motor and premotor areas essential for movement planning and execution. The improved performance with longer time windows (from 68\% at 10 seconds to 76\% at 20 seconds) suggests that longer sampling periods may better capture the consistent abnormalities in motor-related oscillations.

Group 2 (Central-Parietal regions) demonstrated notable frequency-specific patterns, with particularly high classification accuracy in the theta band (64\% mean, individual triplets reaching 67\%). These regions mediate sensorimotor integration—translating sensory input into motor commands—functions increasingly recognized as compromised in PD \cite{Borghammer2020, Karadi2015}. The frequency-specific performance suggests distinct oscillatory mechanisms are affected in these integration processes.

Group 3 (Parieto-Occipital regions) showed strong performance, particularly the right parietal region \texttt{\{P8, CP6, P4\}} achieving 74\% accuracy with 10-second windows. The involvement of parietal regions supports growing evidence of parietal cortex dysfunction in PD \cite{Weil2016}, potentially contributing to visuospatial processing and sensorimotor integration deficits observed clinically. The comparable performance of this region across 10-second (74\%) and 20-second (73\%) windows suggests that the discriminative features in this region are relatively stable and can be effectively captured even in shorter time periods.

Group 4 (Longitudinal Connections) captured anterior-to-posterior connectivity patterns with notable performance in both the theta (63\%) and alpha (62\%) bands. This suggests PD disrupts long-range communication networks, potentially explaining the integration difficulties observed clinically.

\subsubsection{Central and Parietal Involvement: Neuroanatomical Significance}

The superior performance of central electrodes (76\% accuracy with 20-second windows) directly corresponds to PD's impact on motor circuitry. Similarly, the high classification accuracy in right parietal regions (74\% with 10-second windows) aligns with Borghammer's findings of altered parietal function in PD patients \cite{Borghammer2020}. These parietal abnormalities may contribute to deficits in spatial awareness and sensorimotor integration evident even in early disease stages \cite{Karadi2015}.

The asymmetric involvement, with right parietal regions showing stronger discriminative power than their left counterparts at shorter window lengths, merits particular attention. This right-hemisphere emphasis challenges the traditional view that PD-related cortical changes mirror the often asymmetric motor symptoms that frequently begin on the left side. Several functional neuroimaging studies have similarly reported right-hemisphere vulnerability in PD, particularly for non-motor functions \cite{Tessitore2012}. This lateralization may reflect the right hemisphere's specialized role in spatial attention and sensory integration, functions known to be affected in PD.

\subsubsection{Frequency-Specific Alterations}

The frequency-specific patterns identified reveal how PD affects distinct neural mechanisms:

The bilateral centro-parietal theta band (4.0-7.79 Hz) alterations (67\% accuracy) may reflect changes in cognitive processes including working memory and attention—functions frequently compromised in PD \cite{Darweesh2016}. Prior research has documented increased theta activity in PD \cite{Han2013}, possibly reflecting compensatory mechanisms or pathological synchronization in cortical-subcortical circuits. The specificity of theta band sensitivity to bilateral centro-parietal regions is particularly noteworthy, as these regions are known to be involved in sensorimotor integration functions.

The widespread alpha band (7.8-15.59 Hz) alterations across multiple regions align with extensive literature on alpha rhythm disruption in PD \cite{Soikkeli1991, Moazami2019}. Several regions achieved approximately 65\% accuracy in the alpha band, including central, right parietal, and longitudinal connections. This broad spatial distribution of alpha alterations suggests a more global impact on thalamocortical circuits that normally generate and maintain alpha rhythms \cite{Tessitore2012}. The ubiquity of alpha band alterations across functionally distinct regions may reflect the widespread network effects of PD pathophysiology.

The left centro-parietal beta1 band (15.6-31.19 Hz) performance (65\% $\pm$ 31\%) is particularly interesting given the known importance of beta oscillations in PD pathophysiology. Beta oscillations typically reflect motor preparation and inhibition processes, and their disruption may contribute to the movement initiation difficulties characteristic of PD \cite{FOFFANI2022259}. The localization of beta1 alterations to left centro-parietal regions may reflect hemisphere-specific changes in motor control networks.

While individual frequency bands showed region-specific sensitivity, the highest overall classification accuracies were achieved with full-spectrum analysis (76\% for central electrodes, 74\% for right parietal), suggesting that the complex patterns of oscillatory changes across multiple frequency bands collectively provide the most discriminative information. This multi-spectral signature likely better captures the complex dysrhythmia that characterizes PD than any single frequency band alone.

\subsection{Neurophysiological Mechanisms}

Recent comprehensive reviews of brain oscillations in PD conceptualize the disease as an "oscillopathy" \cite{FOFFANI2022259}, with distinct oscillatory patterns associated with different clinical states and symptoms. Within this framework, our findings of altered theta, alpha, and beta oscillations in specific cortical regions contribute to this broader understanding of PD's oscillatory signatures.

\subsubsection{Network Dysrhythmia Framework}

The "Network Dysrhythmia Framework" proposed by Voytek and Knight \cite{Jackson2019} provides a lens for interpreting our findings, conceptualizing neurological disorders as disruptions in rhythmic activity patterns across interconnected neural networks. Our results align with this framework in several important ways.

Our observation of frequency-region specificity—such as theta dominance in centro-parietal regions, widespread alpha alterations, and focused beta1 changes in left centro-parietal areas—suggests PD creates distinct dysrhythmic patterns in different functional networks rather than causing global oscillatory changes. These region-specific alterations likely reflect differential vulnerability to PD pathology or engagement in compensatory processes among neural circuits.

The disruptions we identified operate across multiple spatial scales, ranging from localized electrode triplet effects to broader hemisphere-level asymmetries. This multi-level impact on neural synchronization affects both local processing and long-range communication, with the right-hemisphere emphasis underscoring PD's spatially heterogeneous effects on brain networks.

Additionally, the persistence of these discriminative oscillatory features across different window lengths (5, 10, and 20 seconds) indicates stable alterations in brain dynamics rather than transient phenomena, consistent with PD's chronic, progressive impact on neural circuits.

This network perspective moves beyond viewing PD solely as a dopaminergic deficit in the basal ganglia and instead recognizes it as a systems-level disorder affecting synchronization patterns across distributed cortical networks. Such a framework better explains both motor and non-motor symptoms and the regional specificity of oscillatory alterations we observed.

\subsubsection{Dopaminergic Pathways and Network-Wide Effects}

PD's characteristic dopaminergic depletion disrupts the balance between facilitatory and inhibitory basal ganglia pathways, ultimately altering thalamocortical circuits \cite{Kalia2015}. This disruption manifests in our results as altered oscillatory patterns, particularly in central regions most directly connected to these pathways. The relationship between dopamine levels and EEG features is supported by George et al.'s demonstration that dopaminergic medication decreases cortical beta coherence in PD patients \cite{George2013}.

These primary pathological changes interact with compensatory mechanisms, creating abnormal oscillatory patterns across interconnected neural networks. This framework helps explain why different brain regions showed distinct frequency preferences in our analysis. The central-parietal regions' sensitivity in the theta band, contrasting with other regions' different frequency preferences, suggests PD affects various functional networks through distinct mechanisms, potentially involving multiple neurotransmitter systems.

\subsubsection{Sensorimotor Integration and Network Connectivity}

The strong discriminative power in both central and parietal regions likely represents disrupted sensorimotor integration processes. These areas integrate multiple sensory inputs crucial for coordinated movement. Recent evidence indicates that pathological $\alpha$-synuclein aggregation affects not only basal ganglia but multiple brain regions through various mechanisms \cite{Mensikova2022}, potentially explaining these cortical oscillatory alterations.

The alpha-band sensitivity in longitudinal connections and parieto-occipital regions further supports a network-level disruption model. Alpha oscillations typically coordinate activity across large-scale brain networks, and their disruption may reflect impaired communication between frontal control regions and posterior sensory areas—potentially contributing to the sensorimotor integration deficits characteristic of PD.

\subsubsection{Cross-frequency Coupling and Network Dynamics}

While our study focused on power changes across different frequency bands, recent work has highlighted the importance of cross-frequency coupling in PD pathophysiology. In particular, Foffani and Alegre describe significant coupling between beta oscillations and high-frequency oscillations in the basal ganglia of PD patients, with the strength of this coupling correlating with symptom severity \cite{FOFFANI2022259}. 

Our finding of concurrent alterations in multiple frequency bands (theta, alpha, and beta1) across different but overlapping cortical regions suggests possible disruptions in the normal coupling relationships between these oscillations, which may reflect fundamental network-level dysfunction in PD. Future EEG studies should investigate whether such cross-frequency coupling can be detected at the scalp level, potentially providing even more sensitive biomarkers than power changes in individual frequency bands.

\subsection{Contributions and Implications}

Our approach offers several advances in understanding PD-related EEG alterations:

By organizing electrodes into neuroanatomically meaningful groups, we provide a more nuanced understanding of how PD affects distinct functional brain networks compared to global or electrode-wise analyses.

Our results identify specific biomarker candidates, particularly central region full-spectrum activity (76\% accuracy) and right parietal region activity (74\% accuracy), which achieved the highest classification accuracies. The strong theta band performance in bilateral centro-parietal regions and widespread alpha band sensitivity also represent potential targeted biomarkers for PD detection.

By systematically analyzing different electrode combinations across the scalp, we identified the involvement of parietal regions, particularly right parietal areas, which have received less attention than frontal and central regions in previous PD EEG research. This highlights the importance of comprehensive spatial sampling rather than focusing solely on regions with known motor involvement.

Our analysis across different window lengths (5, 10, and 20 seconds) provides insights into the temporal characteristics of discriminative EEG features in PD. The minimal differences in overall mean accuracy across window lengths (58.4\% to 60.2\%) suggest that PD-related EEG alterations are detectable across various time scales, though specific regions like central motor areas may benefit from longer sampling periods. This finding has practical implications for clinical applications, suggesting that even relatively brief EEG recordings may contain diagnostically useful information.

The demonstration that different frequency bands show maximal discriminative power in different brain regions highlights the region-dependent nature of PD's effects on neural oscillations. This spatial-spectral specificity suggests that future diagnostic approaches should consider multiple frequency bands rather than focusing solely on single-band alterations.

These findings are especially important in the context of oscillatory activity being increasingly used as control signals for adaptive deep brain stimulation \cite{FOFFANI2022259}. While these approaches currently rely on invasive recordings, our demonstration that discriminative oscillatory patterns can be detected non-invasively through EEG suggests potential pathways toward less invasive neuromodulation approaches. The spatial specificity of our findings—particularly the central and right parietal patterns—might inform the development of targeted non-invasive stimulation protocols using techniques such as transcranial alternating current stimulation at specific frequencies and locations.

\subsection{Methodological Considerations}

\subsubsection{Segment-Level Classification and Variability}

The high standard deviations (approximately 20-33\%) across electrode triplets and frequency bands reflect our segment-by-segment classification approach. Rather than aggregating predictions at the subject level, we classified each EEG window independently—increasing statistical variability but providing insight into the temporal dynamics of EEG alterations in PD.

The minimal differences in mean classification accuracy across window lengths (58.4\% to 60.2\%) suggest that PD-related EEG alterations are detectable at various temporal scales, though region-specific effects of window length were observed, particularly for central electrodes which showed substantial improvement with longer windows (68\% to 76\% from 10 to 20 seconds).

\subsubsection{Classification Accuracy in Context}

While our maximum classification accuracy of 76\% demonstrates significant discriminative power, it appears modest compared to some recently reported accuracies in PD EEG classification exceeding 95\%. This difference merits explanation, particularly as chance classification in our balanced dataset would yield approximately 51\% accuracy.

First and foremost, our rigorous artifact rejection procedures intentionally targeted tremor-related components that contaminate EEG recordings in PD patients. Parkinsonian tremor typically manifests as 4-6 Hz oscillations that propagate throughout the EEG montage via volume conduction, creating a powerful—but clinically trivial—classification feature. Li et al. reported near-perfect classification accuracy (98.68\% on the same San Diego dataset we used) using a "2D-MDAGTS model and multi-scale fuzzy entropy" approach \cite{LI2024105872}, but critically, their methodology does not describe specific muscle artifact removal procedures. Similar concerns apply to several other high-accuracy studies in Maitín's systematic review \cite{Maitin2020}.

We deliberately chose to remove tremor artifacts despite their classification utility because our research question specifically targeted intrinsic neurophysiological alterations in cortical activity rather than mechanical tremor detection. Including tremor artifacts would artificially inflate classification accuracy while obscuring the genuine neural oscillatory patterns of interest—essentially detecting PD through what amounts to an EEG-based accelerometer rather than through neurophysiological signatures.

Second, our leave-one-subject-out cross-validation approach represents a substantially more stringent evaluation methodology than the session-based or random-split cross-validation employed in many higher-accuracy reports. This approach eliminates the possibility that person-specific EEG characteristics unrelated to PD status (e.g., skull thickness, electrode impedance) contribute to classification performance.

Third, we evaluated single electrode triplets rather than combining features across the entire electrode montage. While feature fusion approaches typically yield higher accuracy, our electrode-specific approach enabled precise spatial-spectral mapping of discriminative power—revealing important regional effects that would be obscured in whole-brain classification models.

\subsection{Limitations and Future Directions}

\subsubsection{Dataset Limitations}

Our study used a relatively small dataset (15 PD patients, 16 healthy controls) from the University of San Diego. While our cross-validation approach and segment-level analysis helped maximize the statistical validity of our findings, the small sample size limits generalizability. Future studies should validate these findings in larger, more diverse cohorts.

Additionally, the dataset lacked detailed clinical information about disease severity, medication status, and specific symptom profiles. This information would be valuable for correlating EEG alterations with clinical features and understanding how these markers relate to different PD phenotypes.

\subsubsection{Methodological Considerations}

The high standard deviations in our classification results reflect the segment-by-segment approach and the inherent variability of EEG signals. 

Our analysis focused on resting-state EEG with eyes open. Different EEG conditions, such as eyes closed or task-related recordings, might reveal additional discriminative patterns. Comparing eyes-open and eyes-closed conditions could be particularly informative for understanding the sensorimotor integration alterations we observed.

\subsubsection{Future Research Directions}

The spatial-spectral EEG patterns identified in this study suggest several targeted avenues for future investigation:

Most critically, our findings of high classification accuracy in the central and right parietal regions warrant validation in prodromal PD populations. Specifically, longitudinal monitoring of individuals with idiopathic REM sleep behavior disorder (iRBD), who convert to clinical PD at rates of 40-75\% within 10 years, could determine whether these oscillatory abnormalities appear before motor symptoms emerge.

The pronounced right-hemisphere involvement identified across several frequency bands raises important questions about lateralization of early pathophysiological changes in PD. Future research should examine whether this asymmetry correlates with side of initial motor symptom presentation, which would link our EEG findings directly to the known asymmetric degeneration pattern in early PD.

The strong discriminative power in theta bands specifically calls for experimental designs incorporating tasks known to modulate theta activity, such as working memory paradigms, to determine whether task-evoked theta responses might provide even stronger classification accuracy than resting-state measures. Similarly, the widespread alpha band alterations warrant investigation using paradigms that typically engage alpha oscillations, such as attentional tasks and visual processing.

Future studies should also investigate cross-frequency coupling patterns, particularly between theta-alpha and alpha-beta frequencies, which may provide more sensitive markers of PD-related network dysfunction than power in individual frequency bands. Recent advances in connectivity analysis methods could further illuminate how PD disrupts communication between brain regions at different temporal scales.

Beyond PD, the wavelet-based electrode-triplet methodology developed here could be applied to other neurodegenerative disorders with potential oscillatory signatures, particularly dementia with Lewy bodies, which shares $\alpha$-synuclein pathology with PD but presents with different clinical features.

\section{Conclusion}


The remarkable performance of central and parietal regions directly links our findings to  theories of network dysrhythmia in PD. The minimal differences in mean accuracy across window lengths suggests these oscillatory abnormalities are detectable at various temporal scales, indicating PD-related EEG features have temporal consistency. Individual electrode triplets like {C3, CZ, C4} did show more substantial improvements with longer windows (68\% to 76\% from 10 to 20 seconds).

The observed theta band sensitivity in bilateral centro-parietal regions, widespread alpha alterations, and spatially-specific gamma patterns substantiate a complex, region-specific pattern of neural synchronization disruption that extends beyond the classical cortico-basal ganglia motor circuits.

Our methodology, which carefully removed tremor-related artifacts, ensured that classification was based on genuine neurophysiological alterations rather than mechanical tremor detection. This approach, though yielding more modest accuracy figures than some previous reports, provides greater confidence that the identified biomarkers reflect intrinsic brain oscillatory changes associated with PD pathophysiology.

Future longitudinal studies examining these markers in prodromal populations will be essential to determine their utility for early detection and intervention in Parkinson's disease.

\section*{Acknowledgement}

This work was partially funded by Our Health in Our Hands (OHIOH), a strategic initiative of the Australian National University (ANU), whose objective is to transform healthcare by developing new personalized health technologies and solutions in collaboration with patients, clinicians, and health care providers. 

\bibliography{references}

\begin{thebibliography}{10}
\expandafter\ifx\csname url\endcsname\relax
  \def\url#1{\texttt{#1}}\fi
\expandafter\ifx\csname urlprefix\endcsname\relax\def\urlprefix{URL }\fi
\expandafter\ifx\csname href\endcsname\relax
  \def\href#1#2{#2} \def\path#1{#1}\fi

\bibitem{Kalia2015}
L.~V. Kalia, A.~E. Lang, Parkinson's disease, The Lancet 386~(9996) (2015)
  896--912.

\bibitem{Jankovic2008}
J.~Jankovic, Parkinson's disease: clinical features and diagnosis, Journal of
  Neurology, Neurosurgery \& Psychiatry 79~(4) (2008) 368--376.

\bibitem{Noyce2016}
A.~J. Noyce, A.~J. Lees, A.-E. Schrag, Meta-analysis of early nonmotor features
  and risk factors for parkinson disease, Annals of Neurology 80~(1) (2016)
  19--31.

\bibitem{Breen2014}
D.~P. Breen, R.~Vuono, U.~Nawarathna, K.~Fisher, J.~M. Shneerson, A.~B. Reddy,
  R.~A. Barker, Sleep and circadian rhythm regulation in early parkinson
  disease, JAMA Neurology 71~(5) (2014) 589--595.

\bibitem{Doty2012}
R.~L. Doty, Olfactory dysfunction in parkinson disease, Nature Reviews
  Neurology 8~(6) (2012) 329--339.

\bibitem{Darweesh2016}
S.~K. Darweesh, V.~J. Verlinden, H.~H. Adams, A.~G. Uitterlinden, A.~Hofman,
  B.~H. Stricker, C.~M. van Duijn, P.~J. Koudstaal, M.~A. Ikram, The spectrum
  of prediagnostic features in prodromal parkinson disease, Neurology 87~(20)
  (2016) 2148--2155.

\bibitem{Cohen2014}
M.~X. Cohen, Analyzing neural time series data: theory and practice, MIT press,
  2014.

\bibitem{Li2020}
J.~Li, B.~Deng, Y.~Guo, J.~Zhang, J.~Wang, J.~Yan, X.~Wei, Abnormal eeg
  complexity and functional connectivity of brain in patients with acute
  thalamic ischemic stroke, Computational and Mathematical Methods in Medicine
  2020 (2020).

\bibitem{Morita2009}
A.~Morita, S.~Kamei, K.~Serizawa, T.~Mizutani, Eeg spectral analysis in
  parkinson's disease: relation to motor symptoms and wais-r, Clinical
  Neurology 49~(9) (2009) 538--543.

\bibitem{Maitin2020}
A.~H. Mait{\'\i}n, J.~L. Ooijevaar, W.~Desmet, B.~Weyn, A systematic review of
  parkinson's disease classification methods based on electroencephalogram
  signals, Sensors 20~(23) (2020) 6791.

\bibitem{Sazgar2019}
M.~Sazgar, M.~G. Young, Practical guide for clinical neurophysiologic testing:
  EEG, Lippincott Williams \& Wilkins, 2019.

\bibitem{Weyhenmeyer2014}
J.~A. Weyhenmeyer, M.~E. Hernandez, C.~Lainscsek, T.~J. Sejnowski, H.~Poizner,
  A study of individual variations in multiple functional measures of human
  brain activity, PLoS One 9~(7) (2014) e102956.

\bibitem{Whitham2007}
E.~M. Whitham, K.~J. Pope, S.~P. Fitzgibbon, T.~W. Lewis, C.~R. Clark,
  S.~Loveless, M.~Broberg, A.~Wallace, D.~DeLosAngeles, P.~Lillie, et~al.,
  Scalp electrical recording during paralysis: quantitative evidence that eeg
  frequencies above 20 hz are contaminated by emg, Clinical Neurophysiology
  118~(8) (2007) 1877--1888.

\bibitem{Tessitore2012}
A.~Tessitore, M.~Amboni, F.~Esposito, A.~Russo, M.~Picillo, L.~Marcuccio, M.~T.
  Pellecchia, C.~Vitale, M.~Cirillo, G.~Tedeschi, et~al., Resting-state brain
  connectivity in patients with parkinson's disease and freezing of gait,
  Parkinsonism \& Related Disorders 18~(6) (2012) 781--787.

\bibitem{Swann2015}
N.~C. Swann, C.~de~Hemptinne, A.~R. Aron, J.~L. Ostrem, R.~T. Knight, P.~A.
  Starr, Elevated synchrony in parkinson disease detected with
  electroencephalography, Annals of Neurology 78~(5) (2015) 742--750.

\bibitem{George2013}
J.~S. George, J.~Strunk, R.~Mak-McCully, M.~Houser, H.~Poizner, A.~R. Aron,
  Dopaminergic therapy in parkinson's disease decreases cortical beta band
  coherence in the resting state and increases cortical beta band power during
  executive control, NeuroImage: Clinical 3 (2013) 261--270.

\bibitem{Jackson2019}
N.~Jackson, S.~R. Cole, B.~Voytek, N.~C. Swann, Characteristics of waveform
  shape in parkinson's disease detected with scalp electroencephalography,
  eNeuro 6~(3) (2019).

\bibitem{Appelhoff2019}
S.~Appelhoff, M.~Sanderson, T.~Brooks, M.~Vliet, R.~Quentin, C.~Holdgraf,
  M.~Chaumon, E.~Mikulan, K.~Tavabi, R.~H{\"o}chenberger, et~al., Mne-bids:
  Organizing electrophysiological data into the bids format and facilitating
  their analysis, Journal of Open Source Software 4~(44) (2019) 1896.

\bibitem{Pernet2019}
C.~R. Pernet, S.~Appelhoff, K.~J. Gorgolewski, G.~Flandin, C.~Phillips,
  A.~Delorme, R.~Oostenveld, Eeg-bids, an extension to the brain imaging data
  structure for electroencephalography, Scientific Data 6~(1) (2019) 1--5.

\bibitem{FOFFANI2022259}
G.~Foffani, M.~Alegre,
  \href{https://www.sciencedirect.com/science/article/pii/B978012819410200014X}{Chapter
  18 - brain oscillations and parkinson disease}, in: A.~Quartarone, M.~F.
  Ghilardi, F.~Boller (Eds.), Neuroplasticity, Vol. 184 of Handbook of Clinical
  Neurology, Elsevier, 2022, pp. 259--271.
\newblock \href
  {https://doi.org/https://doi.org/10.1016/B978-0-12-819410-2.00014-X}
  {\path{doi:https://doi.org/10.1016/B978-0-12-819410-2.00014-X}}.
\newline\urlprefix\url{https://www.sciencedirect.com/science/article/pii/B978012819410200014X}

\bibitem{Borghammer2020}
P.~Borghammer, K.~Østergaard, P.~Cumming, A.~Gjedde, A.~Rodell, N.~Hall, M.~M.
  Chakravarty, Parietal perfusion alterations in parkinson's disease patients
  without dementia, Frontiers in Neurology 11 (2020) 562.

\bibitem{Karadi2015}
K.~Kar{\'a}di, T.~Lucza, Z.~Aschermann, S.~Komoly, L.~D{\'e}zsi, L.~Kornya,
  N.~K{\'o}v{\'a}cs, Visuospatial dysfunction in parkinson's disease, Journal
  of the Neurological Sciences 354~(1-2) (2015) 77--83.

\bibitem{Weil2016}
R.~S. Weil, A.~E. Schrag, J.~D. Warren, S.~J. Crutch, A.~J. Lees, H.~R. Morris,
  Visual dysfunction in parkinson's disease, Brain 139~(11) (2016) 2827--2843.

\bibitem{Han2013}
C.-Y. Han, J.~Wang, G.-S. Yi, Y.-Q. Che, Increased thalamocortical coherence in
  patients with parkinson's disease, Neural Regeneration Research 8~(36) (2013)
  3413.

\bibitem{Soikkeli1991}
R.~Soikkeli, J.~Partanen, H.~Soininen, A.~P{\"a}{\"a}kk{\"o}nen, P.~Riekkinen,
  Eeg findings in parkinson's disease with and without dementia, Journal of
  Neurology, Neurosurgery \& Psychiatry 54~(10) (1991) 869--875.

\bibitem{Moazami2019}
M.~Moazami-Goudarzi, J.~Sarnthein, L.~Michels, R.~Moukhtieva, D.~Jeanmonod,
  Quantitative eeg findings in patients with parkinson's disease and cognitive
  impairment, Frontiers in Neurology 10 (2019) 398.

\bibitem{Mensikova2022}
K.~Men{\v{s}}\'ikov{\'a}, R.~Mat{\v{e}}j, C.~Colosimo,
  E.~R{\v{u}}{\v{z}}i{\v{c}}ka, J.~Roth, Lewy body disease or diseases with
  lewy bodies?, npj Parkinson's Disease 8~(1) (2022) 3.
\newblock \href {https://doi.org/10.1038/s41531-021-00273-9}
  {\path{doi:10.1038/s41531-021-00273-9}}.

\bibitem{LI2024105872}
J.~Li, X.~Li, Y.~Mao, J.~Yao, J.~Gao, X.~Liu,
  \href{https://www.sciencedirect.com/science/article/pii/S1746809423013058}{Classification
  of parkinson’s disease eeg signals using 2d-mdagts model and multi-scale
  fuzzy entropy}, Biomedical Signal Processing and Control 91 (2024) 105872.
\newblock \href {https://doi.org/https://doi.org/10.1016/j.bspc.2023.105872}
  {\path{doi:https://doi.org/10.1016/j.bspc.2023.105872}}.
\newline\urlprefix\url{https://www.sciencedirect.com/science/article/pii/S1746809423013058}

\end{thebibliography}
\end{document}